\documentclass[twocolumn,showpacs,preprintnumbers,amsmath,amssymb,prl]{revtex4}
\usepackage{graphicx}
\usepackage{color}
\begin{document}
\title{Anisotropically high entanglement of biphotons generated in spontaneous parametric down
conversion}
\author{M.V. Fedorov, M.A. Efremov, P.A. Volkov }
\affiliation{A.M.~Prokhorov General Physics Institute of Russian
Academy of Sciences, Russia}
\author{E. V. Moreva}
\email{ekaterina.moreva@gmail.com}\affiliation{Moscow Engineering
Physics Institute (State University), Russia}
\author{S.S. Straupe, S.P. Kulik } \affiliation{Faculty of Physics, Moscow State
University, Russia}

\date{\today}

\begin{abstract}

We show that the wave packet of a biphoton generated via spontaneous
parametric down conversion is strongly anisotropic. Its anisotropic
features manifest themselves very clearly in comparison of
measurements performed in two different schemes: when the detector
scanning plane is perpendicular or parallel to the plane containing
the crystal optical axis and the laser axis. The first of these two
schemes is traditional whereas the second one gives rise to such
unexpected new results as anomalously strong narrowing of the
biphoton wave packet measured in the coincidence scheme and very
high degree of entanglement. The results are predicted theoretically
and confirmed experimentally.

\end{abstract}


 \maketitle

 \section{Introduction}

Quantum entanglement is one of surprising consequences of quantum
mechanics. It lies in the center of attention since the famous paper
by Einstein, Podolsky, and Rosen\cite{EPR} (EPR). Two (or more)
subsystems are entangled if the system as a whole is characterized
by a wave function or a density matrix, which cannot be presented in
the form of products of subsystems' wave functions or density
matrices \cite{Schrodinger}. In some cases entanglement (for example
Bell states) leads to complete determinacy of the combined bipartite
system while states of the subsystems are fully undetermined.
Entanglement is the key element of such phenomena and domains of
modern quantum optics as quantum teleportation, quantum
cryptography, Bell violation experiments, quantum computation.

A special class of entangled systems, attracting a permanently
growing attention, is that of systems with continuous variables.
The most often and widely studied example is Spontaneous
Parametric Down Conversion (SPDC) \cite{Rubin, Klyshko, Monken1,
Monken2, Eberly-Law, Howell, JPB, Dangelo}. Some questions arising
in discussions of entanglement in such systems are how high is the
entanglement in a given multipartite state, what is the best
measure of the degree of entanglement, how the degree of
entanglement can be measured experimentally, etc., etc. It's
rather well recognized \cite{Chan-Eberly-Law, Eberly-Law} that in
the case of pure bipartite states the degree of their entanglement
can be characterized by the so called Schmidt number $K$
\cite{Grobe, Knight}. However, the Schmidt number definition does
not provide any recipes on how it could be measured. An
alternative approach \cite{JPB, PRA} to the entanglement
characterization is based on analysis of conditional and
unconditional probability density distributions, which are related
directly to experimentally measurable coincidence and
single-particle wave packets of particles. Specifically, it was
suggested \cite{PRA} to characterize the degree of entanglement by
the ratios $R_x=\Delta x^{(s)}/\Delta x^{(c)}$ or $R_k=\Delta
k^{(s)}/\Delta k^{(c)}$, where $\Delta x^{(s,\,c)}$ and $\Delta
k^{(s,\,c)}$ are the single-particle $(s)$ and coincidence $(c)$
coordinate and momentum wave packet widths, $x$ and $k$ are the
coordinate and momentum variables of one of two particles under
consideration. As it was shown \cite{JPB}, $K\equiv R_x\equiv R_k$
for a rater general class of double-Gaussian bipartite wave
functions
\begin{equation}
 \label{double-Gaussian}
 \exp\left\{-\frac{(\alpha x_1+\beta
 x_2)^2}{2a^2}\right\}\times\exp\left\{-\frac{(\gamma x_1+\delta
 x_2)^2}{2b^2}\right\},
\end{equation}
where $x_1$ and $x_2$ are continuous variables of two particles,
$\alpha$, $\beta$, $\gamma$, $\delta$ and $a$, $b$ are constants.
Physical meaning of these parameters is determined by a problem
under consideration. There are problems in which a double-Gaussian
bipartite wave function  arises quite naturally. One example of such
a phenomenon is the strong-field electron-positron pair production
\cite{OC}. In some other processes and phenomena the arising
bipartite wave functions are not double-Gaussian but can be
successfully modeled by double-Gaussian functions with appropriately
chosen parameters. In these cases the parameters $K$, $R_x$, and
$R_k$ are usually close if not equal to each other. And, finally, if
a bipartite wave function has a form strongly different from the
double-Gaussian one, all three parameters $K$, $R_x$, and $R_k$ can
differ from each other. We assume that in such a case each of these
parameters characterizes the degree of entanglement to be seen in
different ways, and entanglement itself cannot be characterized
completely by any single universal parameter.

Note that for non-entangled states $K=R_x=R_k=1$. Moreover, for
non-entangled states coincidence and single wave packets are
identical to each other. Any differences between the coincidence and
single wave packets can be considered as a manifestation of
entanglement. Macroscopic characteristics of these differences are
given just by the parameters $R_x$, $R_k$, and $K$.  If these
parameters differ from each other, the largest of them indicates an
optimal way of observing entanglement experimentally or calculating
mathematically, whereas in other ways of measurements or
calculations entanglement can be hidden.

Also, one can characterize the degree of entanglement by the
EPR-related parameter given by the inverse coordinate and momentum
conditional uncertainties or wave packet widths for any given
particle: $C_{\rm EPR}=1/[\Delta x^{(c)}\times\Delta k^{(c)}]$
\cite{Howell, JPB, PRA}. For non-entangled states a specific value
of this parameter depends on the definition of widths and on the
shape of single-particle wave functions $\psi_1(x_1)$ and
$\psi_2(x_2)$, the product of which gives the non-entangled
bipartite wave function. If the widths are determined as square
roots of variances (as in the work \cite{Howell}), and the functions
$\psi_1(x_1)$ and $\psi_2(x_2)$ are Gaussian, the no-entanglement
value of the EPR-related parameter is given by $C_{{\rm EPR}\;{\rm
var}}^{(0)}=2$. If the widths $\Delta k^{(c)}$ and $\Delta x^{(c)}$
are determined as the widths  of the corresponding probability
distribution curves at the half-maximum level (as in this work), for
the same forms of $\psi_1$ and $\psi_2$, $C_{{\rm EPR}\;
1/2}^{(0)}=(4\ln 2)^{-1}\approx 0.36$. In terms of the EPR-related
parameter $C_{\rm EPR}$, the degree of entanglement for
non-factorized bipartite functions is evaluated as $C_{\rm
EPR}/C_{\rm EPR}^{(0)}$.

It is assumed usually that deviations of realistic bipartite wave
functions from the double-Gaussian one [Eq.
(\ref{double-Gaussian})] do not give rise to any pronounced new
features of the processes under consideration and, in particular,
all the entanglement parameters remain of the same order of
magnitude if not equal identically. In contrast to this
assumption, we show below that, in the case of SPDC, deviations of
the biphoton wave function (to be found below) from the
double-Gaussian one are well pronounced and they can change
drastically the observable wave packet pictures. Owing to this,
the SPDC entanglement phenomenon appears to be essentially
multiparametric, with the above-indicated parameters significantly
differing from each other.

Concerning the observable biphoton wave packet pictures,
formulated explicitly or not, the usual notion is that they are
more or less invariant with respect to rotations around the pump
laser axis. In contrast to this, we show that the SPDC biphotons
are strongly anisotropic and the degree of anisotropy depends
essentially on relative orientation of a crystal and the
observation direction. Furthermore, usually the biphoton momentum
wave packet structure is supposed to be controlled by the laser
pump angular distribution and by geometrical factors like crystal
shape etc. \cite{Klyshko, Monken1, Monken2}. That is why it seems
that in order to generate well-localized biphoton wave packets in
the $k$-domain one has to use good laser sources with a very small
angular divergency and/or long crystals. At the same time forming
particular angular profile of the pump is needed for coupling
biphotons into the optical fibers \cite{Kurtsiefer, Bennik}, that
is very important for some applications. So the question arises
whether it's possible to get a good biphoton wave packet
localization with a poor laser having a relatively high angular
divergence? Is it possible to vary the degree of entanglement
while keeping constant parameters of the pump and nonlinear
crystal generating photon pairs? The answer we give in this paper
is yes, owing to the above-described anisotropy, it's possible to
control the degree of wave packet localization and the degree of
entanglement simply by tuning a crystal orientation with respect
to detectors position. Under the optimal conditions the achievable
degrees of wave packet localization and entanglement appear to be
very high. In particular, it becomes possible to get the
coincidence angular biphoton distribution much narrower than that
of the pump, which can look rather unexpected and unusual from the
point of view of traditional theoretical and experimental
approaches to the investigation of SPDC entanglement. The results
outlined above follow from the theoretical formulas to be derived
below and they are confirmed by direct experimental observations.

\section{Theoretical description}
\subsection{Derivation of the main general formula}
\begin{figure}[h]
\centering\includegraphics[width=8cm]{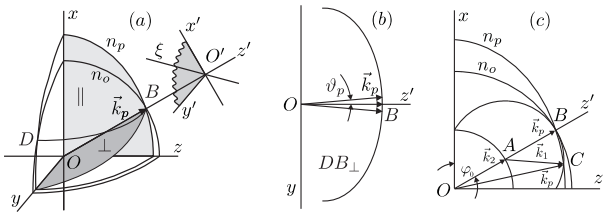}
\caption{{\protect\footnotesize {$(a)$ Octant of the refractive
index surfaces $n_e({\vec r})$ and $n_o({\vec r})$ for pump and
signal photons, $(b,c)$ are two perpendicular section by the planes
$\perp$ and $\|$ shaded in $(a)$; $DB_\perp$ is the projection on
the $\perp$-plane of the circle shown partially by the arc $DB$ in
$(a)$; $\vartheta_p$ is the angle between ${\vec k}_p$ and the
$z^\prime$ axis.}}} \label{Fig1}
\end{figure}

Let us consider a collinear and degenerate type I SPDC process, when
extraordinary pump photon of a frequency $\omega_p$ decays in two
ordinary photons (signal and idler) with equal frequencies
$\omega_p/2$ and propagating more or less along the pump beam. In
3D, refractive index surfaces for extraordinary [$n_e({\vec
r};\,\omega_p)$] and ordinary [$n_o({\vec r};\,\omega_p/2)$] waves
in an anisotropic crystal are, correspondingly, an ellipsoid and a
sphere. Fig. \ref{Fig1}($a$) shows sections of these figures by
three coordinate planes $(xz)$, $(xy)$, and $(yz)$. The optical axis
of a crystal is taken directed along
the $Ox$ axis, and the orthogonal axes $Oy$ and $Oz$ are chosen in
such a way that the $(xz)$ plane contains the pump laser axis
$Oz^\prime$. The arc $DB$ in Fig. \ref{Fig1}$(a)$ is a part of the
circle by which the sphere and ellipsoid cross each other. We assume
that the laser axis is directed strictly to the point $B$ where the
arc $DB$ crosses the $xz$ plane. This means that for pump and
emitted photons propagating strictly along the $z^\prime$-axis the
collinear degenerate phase matching condition is exactly satisfied,
${\vec k}_p={\vec k}_1+{\vec k}_2$ at ${\vec k}_p\,\|\,{\vec
k}_1\,\|\,{\vec k}_2\,\|\,Oz^\prime$, $k_1=k_2=k_p/2$. In our
consideration we assume that the pump is not a single plane wave but
is given by a coherent superposition of plane waves with wave
vectors ${\vec k}_p$ filling in a cone, the axis of which coincides
with the $Oz^\prime$ axis, and the angular width $\alpha$ is finite.

$O^\prime$ in Fig. \ref{Fig1}($a$) is some point in a far zone
located at the laser axis $Oz^\prime$ and such that detectors
registering photons are located in the plane perpendicular to
$Oz^\prime$ and containing the point $O^\prime$. The axes $O^\prime
x^\prime$ and $O^\prime y^\prime$ in this plane are perpendicular to
each other and to $O^\prime z^\prime$, and $O^\prime x^\prime\in
(xz)$, $O^\prime y^\prime\|Oy$. Let us assume that detectors are
installed along some line $O^\prime\xi$ in the plane $(x^\prime
y^\prime)$ and, hence, they register only photons (1) and (2) with
wave vectors ${\vec k}_1$ and ${\vec k}_2$ belonging to the plane
($z^\prime\xi$). Orientation of the detector-installation line
$O^\prime\xi$ in the plane ($x^\prime y^\prime$) can be changed with
two limiting cases given by $O^\prime\xi=O^\prime y^\prime\|Oy$ and
$O^\prime\xi=O^\prime x^\prime\in (xz)$ . In these two limiting
cases the detectors register only photons with wave vector ${\vec
k}_1$ and ${\vec k}_2$ belonging, correspondingly, to the
$yOz^\prime$ and $xOz$ planes. In Fig. \ref{Fig1}$(a)$ these planes
are shaded and they are labeled with the symbols $\perp$ and $\|$,
respectively. Here ``perpendicular$"$ and ``parallel$"$ mean that
the observation plane (i.e., the plane containing wave vectors of
photons to be observed) is perpendicular or parallel to the plane
containing the crystal optical axis and the laser axis. Below these
two cases are referred to as those of the perpendicular and parallel
geometry. Sections of the $3D$ refractive index surfaces by the
$\perp$ and $\|$ planes are shown in the pictures $(b)$ and $(c)$ of
Fig. \ref{Fig1}.

Note that the idea of changing orientation of the
detector-installation line is used here for simplification of
illustrations like those given in Fig. \ref{Fig1}. In real
experiment the laser pump axis was horizontal ($\|\,Oz$), and
detectors were installed in such a way that they registered only
photons with horizontally oriented wave vectors ${\vec k}_1$ and
${\vec k}_2$. This was a crystal that was rotated around the laser
axis (together with the laser polarization) instead of changing
detector positions. There are the following rules of one-to-one
correspondence between these experimental conditions and the
above-described geometries of Fig. \ref{Fig1}: (1) orientation of
the crystal optical axis in the horizontal plane $(yz)$ is
equivalent to the $\|$  geometry and (2) orientation of the
crystal optical axis in the vertical plane $(xz)$ is equivalent to
the  $\perp$ geometry of Fig. \ref{Fig1}.

The second note concerns an additional assumption to be used below
for simplicity and not motivated by orientation reasons. In terms of
notations used in Fig. \ref{Fig1}$(a)$, let us assume that the
photons (1) and (2) with wave vectors ${\vec k}_1$ and ${\vec k}_2$
belonging to the plane $(z^\prime\xi)$ arise only owing to the decay
of pump photons with wave vectors ${\vec k}_p$ belonging to the same
plane $(z^\prime\xi)$. Rigorously, this is not necessarily true:
owing to the non-collinear phase matching processes photons (1) or
(2) can be emitted with wave vectors ${\vec k}_1$ or ${\vec k}_2$
belonging to the plane $(z^\prime\xi)$ even if the pump wave vector
${\vec k}_p$ is oriented differently. Such processes can affect the
single-particle momentum distributions of emitted photons, and this
effect is discussed below. But, on the other hand, the non-collinear
phase matching processes can be effectively suppressed by means of
axially asymmetric focusing of a laser beam. This possibility is
also discussed below and its experimental realization is presented.

Under the formulated assumptions, with the help of the results
derived by Monken et al in 1998 \cite{Monken1}, in the
approximation of a wide crystal we can write down immediately the
following expression for the biphoton wave function depending on
the transverse components of the emitted photon momenta ${\vec
k}_{1\,\xi}$ and ${\vec k}_{2\,\xi}$

\begin{eqnarray}
\Psi(k_{1\,\xi},\,k_{2\,\xi})\propto
E_p^*(k_{1\,\xi}+k_{2\,\xi})\times\mathrm{sinc}\left(\frac{L\,\Delta_{z^\prime}}{2}\right),
\label{wide-crystal}
\end{eqnarray}
where ${\rm sinc}(u)=\sin (u)/u$, $L$ is the crystal length in the
$z^\prime$-direction, and $\Delta_{z^\prime}$ is the longitudinal
detuning
\begin{eqnarray}
 \notag
 \Delta_{z^\prime}=k_{p\;z^\prime}-k_{1\,z^\prime}-k_{2\,z^\prime}\quad\quad\quad\\
 =\sqrt{k_p^2-k_{p\,\xi}^2}-\sqrt{k_1^2-k_{1\,\xi}^2}-\sqrt{k_2^2-k_{2\,\xi}^2}.
 \label{detuning}
\end{eqnarray}

The above-mentioned approximation of a wide crystal means that its
size in the direction parallel to the $O^\prime\xi$ axis is large
enough to make the transverse momentum conservation rule, or
transverse phase matching condition, exactly satisfied to give
$k_{p\,\xi}=k_{1\,\xi}+k_{2\,\xi}$. This relation is used in the
definition of the momentum-dependent pump field-strength amplitude
in Eq. (\ref{wide-crystal}),
$E_p^*(k_{p\,\xi})=E_p^*(k_{1\,\xi}+k_{2\,\xi})$, and it can be used
also in the definition of the longitudinal detuning
$\Delta_{z^\prime}$ (\ref{detuning}).

To simplify further Eq. (\ref{detuning}) for the longitudinal
detuning we can use the near-axis approximation in which
$|k_{1,2\,\xi}|\ll k_{1,2}$, $k_{p\,\xi}\ll k_p$. By expanding
square  roots in Eq. (\ref{detuning}) in powers of transverse
components of all the wave vectors and keeping only two first
orders we get
\begin{equation}
 \label{near-axis}
 \Delta_{z^\prime}=k_p-k_1-k_2+\frac{(k_{1\,\xi}-k_{2\,\xi})^2}{2k_p}\,,
\end{equation}
where in the second term on the right-hand side we took $k_p\approx
2k_1=2k_2$. As for the first term, $k_p-k_1-k_2$, if it can be taken
equal zero identically, then Eqs. (\ref{wide-crystal}) and
(\ref{near-axis}) give immediately the well-known and widely used
formula of the work \cite{Monken2}
\begin{eqnarray}
\Psi(k_{1\,\xi},\,k_{2\,\xi})\propto
E_p^*(k_{1\,\xi}+k_{2\,\xi})\mathrm{sinc}\left(\frac{L\,(k_{1\,\xi}-k_{2\,\xi})^2}{4k_p}\right).
\label{Monken2}
\end{eqnarray}
This is the formula that is most often met and used for theoretical
description of the SPDC biphoton entanglement. Eq. (\ref{Monken2})
has a form very close to that of the double-Gaussian wave function
given by Eq. (\ref{double-Gaussian}). Indeed, very often the pump
envelope has a Gaussian form. As for the sinc$^2$-function, it can
be rather successfully modeled by a Gaussian function too
\cite{Eberly-Law}. The key point of such a reduction to the double
Gaussian form is the separation of variables $k_{1\,\xi}$ and
$k_{2\,\xi}$ for their two independent linear combinations
$k_{1\,\xi}+k_{2\,\xi}$ and $k_{1\,\xi}-k_{2\,\xi}$ in the arguments
of two protofunctions of Eq. (\ref{Monken2}), $E_p^*$ and sinc. As
we will show now, such separation of variables and reduction of the
wave function to the double-Gaussian form essentially depend on the
assumption $k_p-k_1-k_2\equiv 0$ which is true only under very
special conditions.

As it can be seen from the pictures of Fig. \ref{Fig1}, the
collinear phase matching condition $k_p-k_1-k_2= 0$ can be taken
satisfied for different orientations of the pump wave vector ${\vec
k}_p$ only if the detection direction $O^\prime\xi$ is perpendicular
to the plane containing the laser and crystal optical axes, i.e., if
$O^\prime\xi\|Oy$. In this case, for all wave vectors ${\vec k}_p$
located in the $\perp$-plane of Fig. \ref{Fig1}$(a)$ and not too far
deviating from the $Oz^\prime$ axis, their length can be taken
approximately equal: $k_p(\vartheta_p)\approx k_p(0)=k_1+k_2$, where
$\vartheta_p$ is the angle between ${\vec k}_p$ and $Oz^\prime$,
$|\vartheta|\ll 1$. Owing to the last condition the arc $DB_\perp$
in Fig. {\ref{Fig1}}$(b)$ can be approximated by an arc of a circle
with the center at the origin $O$, and then equality of all wave
vector lengths $k_p(\vartheta)$ becomes evident.

At all other orientations of the detector-installation line
$O^\prime\xi$, non-parallel to $Oy$ direction, the difference
$k_p(\vartheta_p)-k_1-k_2$ on the right-hand side of Eq.
(\ref{near-axis}) cannot be taken equal zero at $\vartheta_p\neq 0$
even approximately. At small values of $\vartheta_p$ the dependence
$k_p(\vartheta_p)$ can be linearized to give
\begin{equation}
k_p(\vartheta_p)=\frac{\omega}{c}\,n_p(\varphi_0+\vartheta_p)\approx
k_p(0)\left[1+\vartheta_p\frac{n_p^\prime(\varphi)}{n_p(\varphi)}\right]_{\varphi=\varphi_0}\,,
\label{kp(theta-p)}
\end{equation}
where $n_p^\prime(\varphi)=dn_p(\varphi)/d\varphi$, $\varphi$ is the
angle between ${\vec k}_p$ and the crystal optical axis ($Ox$ in
Fig. \ref{Fig1}) and $\varphi_0$ is the angle between the optical
and laser axes. On the other hand, as the laser axis $Oz^\prime$ and
the detector-installation line $O^\prime\xi$ are perpendicular to
each other, at small $\vartheta_p$ we have
$k_p(0)\,\vartheta_p\approx k_{p\,\xi}=k_{1\,\xi}+k_{2\,\xi}$. By
remembering that $k_p(\vartheta_p=0)=k_1+k_2$ and combining all
these relations together we get the following expression for the
longitudinal detuning $\Delta_{z^\prime}$
\begin{equation}
 \label{detunng2}
 \Delta_{z^\prime}=\frac{n_p^\prime}{n_o}(k_{1\,\xi}+k_{2\,\xi})
 +\frac{(k_{1\,\xi}-k_{2\,\xi})^2}{2k_p}
\end{equation}
and for the momentum biphoton wave function
\begin{eqnarray}
\Psi(k_{1\,\xi},\,k_{2\,xi})\propto
E_p^*(k_{1\,\xi}+k_{2\,\xi})\notag\quad\quad\quad\\\times\,\mathrm{sinc}\left\{\
\frac{L}{2}\left[\frac{n_p^\prime}{n_o} (k_{1\,\xi}+k_{2\,\xi})+
\frac{(k_{1\,\xi}-k_{2\,\xi})^2}{2k_p}\right]\right\}\,,
\label{New-formula}
\end{eqnarray}
where $n_p^\prime=n_p^\prime(\varphi_0)$ and $n_o=n_p(\varphi_0)$.
Eq. (\ref{New-formula}) is the main new formula for the biphoton
wave function we suggest instead of the traditional one given by Eq.
(\ref{Monken2}). These two formulas differ from each other  by the
first term in the sum in square brackets under the symbol of the
${\rm sinc}-$function in Eq. (\ref{New-formula}). Below we will
analyze how important this addition is. But before doing this, let
us rewrite Eq. (\ref{New-formula}) in terms of the scattering angles
$\theta_1$ and $\theta_2$, defined outside the crystal as
$\theta_{1,\,2}=2k_{1,\,2\;\xi}/k_p^{(0)}$, where
$k_p^{(0)}=\omega_p/c$. By using these definitions we express the
transverse components of the emitted photon wave vectors via the
scattering angles and reduce Eq.  (\ref{New-formula}) to the form
\begin{eqnarray}
 \Psi(\theta_1,\,\theta_2)\propto
 {\widetilde E}_p^*\left(\frac{\theta_1+\theta_2}{2}\right)\quad\quad\quad\notag\\
 \label{new-via-theta}
 \times\,{\rm sinc}\left\{\displaystyle\frac{Lk_p^{(0)}}{16n_o}
 \Big[4n_p^\prime(\theta_1+\theta_2)+(\theta_1-\theta_2)^2\Big]\right\}\,,
\end{eqnarray}
where ${\widetilde E}_p(\theta_p)$ is the pump amplitude angular
distribution outside the crystal,
$\theta_p=\frac{1}{2}(\theta_1+\theta_2)=n_o\vartheta_p$ (to remind,
by definition $\vartheta_p$ is the angle between ${\vec k}_p$  and
the laser axis $Oz^\prime$ in a crystal) . In terms of scattering
angles the longitudinal detuning of Eq. (\ref{detunng2}) appears to
be given by
\begin{equation}
 \label{detuning-3}
 \Delta_{z^\prime}=\frac{k_p^{(0)}}{8}\Big[4n_p^\prime(\theta_1+\theta_2)+(\theta_1-\theta_2)^2\Big].
\end{equation}

\begin{figure}[h]
\centering\includegraphics[width=8.5cm]{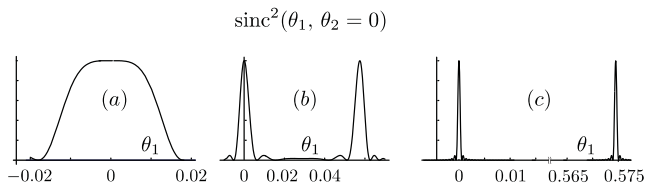}
\caption{{\protect\footnotesize {The function ${\rm sinc}^2
(L\Delta_{z^\prime}/2)$ at $\theta_2=0$ and $(a)\,n_p^\prime=
\,0,\,(b)\,-0.01436,\,\mathrm{and} \,(c) -0.1436$; $\theta_1$ is
in radians.}}} \label{Fig2}
\end{figure}

The new first term in the square brackets of Eqs.
(\ref{new-via-theta}) and (\ref{detuning-3}) is linear whereas the
second term is quadratic in small angles $\theta_1$ and $\theta_2$.
For this reason we can expect that the linear term is even more
important than the quadratic one, if only the refractive index
derivative $n_p^\prime$ is not negligibly small. But this does not
mean that the quadratic term can be dropped because in such an
approximation we would get infinitely wide single-particle
distributions. So, both linear and quadratic terms have to be taken
into account. The role of the linear term is illustrated by three
curves of Fig. \ref{Fig2} where
the function ${\rm sinc}^2(L\Delta_{z^\prime}/2)$ is plotted in its
dependence on $\theta_1$ at $\theta_2=0$ and at three different
values of the refractive index derivative. The curves of Fig.
\ref{Fig2} are calculated for the pump and crystal parameters
corresponding to the experimental ones (see the following Section):
the pump wavelength $\lambda_p=325$ nm and LiIO$_3$ crystal of a
length $L=1.5\,{\rm}$ cm. In this case we have found
$\varphi_0\approx 60^{\rm o}$ and $n_p^\prime=-0.1436$ for the
detection of photons in the vertical plane of Fig. \ref{Fig1}. The
three curves of Fig. \ref{Fig2} correspond to $(a)\, n_p^\prime=0$ -
photon propagation and detection in the horizontal plane,
$(b)\,n_p^\prime=-0.01436$ - some intermediate case between $\perp$-
and $\|$-geometries of Fig. \ref{Fig1}$(a)$, and $(c)\,
n_p^\prime=-0.1436$ - propagation and detection in the $\|$-plane.
One can see, that with a growing value of $|n_p^\prime|$ the
structure of the curves in Fig. \ref{Fig2} changes drastically. A
single wide peak splits for two peaks, spacing between them grows,
and the peaks themselves are getting very narrow. The widths of the
only peak at $n_p^\prime=0$ and of narrow peaks at
$n_p^\prime=-0.1436$ are equal to 24 mrad and 0.5 mrad,
correspondingly. The ratio of these two widths equals to 48 and this
number characterizes the degree of peak narrowing arising owing to
modifications in the formula of Eqs. (\ref{New-formula}),
(\ref{new-via-theta})  compared to that of Eq. (\ref{Monken2}).

Mathematically splitting of a single peak for two ones follows from
the quadratic dependence of the detuning $\Delta_{z^\prime}$
(\ref{detuning-3}) on $\theta_1$. The ${\rm sinc}-$function is
maximal at zero value of its argument, i.e., at zero detuning
$\Delta_{z^\prime}$. In a  general case the quadratic equation
$\Delta_{z^\prime}(\theta_1)=0$ has two solutions which correspond
to two peaks in the dependence of sinc$^2$ on $\theta_1$. The only
exception is the degenerate case $n_p^\prime=0$ when two solutions
of the quadratic equation merge into one. Qualitatively the same
conclusions and the same arguments are illustrated by the Fig.
\ref{Fig1}$(c)$. In this picture the vector ${\vec k}_2$ is plotted
along the optical axis $Oz^\prime$ ($\theta_2=0$), and it ends at
the point $A$. The ending locus for vectors ${\vec k}_1$ is given in
this case by a circle of the radius $n_0/2$ (in units of
$\omega_p/c)$ and with the center at the point $A$. As it's seen
from the picture, there are two points $B$ an $C$ where this circle
crosses the ellipse $n_p(\theta_p)$ which, in its turn, is the
ending locus for the pump wave vectors ${\vec k}_p$. This picture
shows that in the case $n_p^\prime\neq 0$ there are two directions
of the vectors ${\vec k}_1$ and ${\vec k}_p$ (at a given ${\vec
k}_2$) in which the exact phase matching condition ${\vec k}_p={\vec
k}_1+{\vec k}_2$ appears to be fulfilled. Quantitatively, the
nonzero angle of the second exact phase matching direction appears
to be very large. The value of $\theta_1\approx 0.57$ in Fig.
\ref{Fig2}$(c)$ corresponds to about 30$^{\rm o}$ for the scattering
angle $\theta_1$ and to $\theta_p\approx\,15^{\rm o}$ for the pump.
As the last value is much larger than the pump angular divergence
(typically $\alpha\sim 4\,{\rm mrad}\approx 0.23^{\rm o}$), the
second (nonzero-angle) peak of the sinc$^2$ function does not give
contributions to the single-particle photon momentum distributions
(see the derivation and discussion below). The large-angle second
peak of the sinc$^2$-function was not observed also in the
coincidence measurements described in the following Section simply
because this was out of the detector scanning range. For these
reasons we restrict our further analysis by considering only one
peak of the sinc$^2$-function located at small values of the
scattering angle $\theta_1$ in both cases of zero and nonzero values
of the refractive index derivative $n_p^\prime$.

The quadratic dependence of $\Delta_{z^\prime}$ on $\theta_1$ and a
transition from the case when the equation $\Delta_{z^\prime}=0$ has
a single solution to the case of two different solutions (at
$n_p^\prime\neq 0$) explain also qualitatively the reasons of peak
narrowing. Roughly the sinc$^2$ peak widths can be evaluated from
the condition $|\Delta_{z^\prime}|\sim 1/L$. In the case
$n_p^\prime=0$ at $\theta_2=0$ we have $\Delta_{z^\prime}\sim
k_p^{(0)}\theta_1^2$ which gives $\Delta\theta_1\sim
1/\sqrt{Lk_p^{(0)}}$. In contrast, in the case $n_p^\prime\neq 0$ in
a vicinity of each narrow peak the detuning is approximately linear
in $\theta_1$, and from the same condition $|\Delta_{z^\prime}|\sim
1/L$ we find $\Delta\theta_1\sim 1/(Lk_p^{(0)})$. As the product
$Lk_p^{(0)}$ is rather large ($\sim 10^5$), both width are small but
the peak widths occurring in the case $n_p^\prime\neq 0$ is much
smaller than that occurring in the case $n_p^\prime=0$,
$1/(Lk_p^{(0)})\ll 1/\sqrt{Lk_p^{(0)}}$ .

\subsection{Coincidence and single-particle distributions}

Coincidence distributions of photons are given simply by the
squared absolute value of the wave function
$\Psi(\theta_1,\,\theta_2)$ of Eq. ({\ref{new-via-theta}}) at a
given value of one of the angles, $\theta_1$ or $\theta_2$. E.g.,
at $\theta_2=0$
\begin{eqnarray}
\frac{dw^{(c)}(\theta_1)}{d\theta_1}\propto
\left|\Psi\left(\theta_1,\,0\right)\right|^2
\notag\quad\quad\quad\\\propto \left|{\widetilde
E}_p\left(\frac{\theta_1}{2}\right)\, {\rm
sinc}\left[\frac{Lk_p}{16}\,\left(4n_p^\prime\,\theta_1+\theta_1^2\right)\right]\right|^2.
\label{coinc}
\end{eqnarray}

Single-particle distributions are given by the squared wave
function of Eq. (\ref{new-via-theta}) integrated over, e.g.,
$\theta_2$. In the case $n_p^\prime\neq 0$ it is convenient to
make the integration variable substitution
$\theta_2\rightarrow\theta\equiv\theta_2+\theta_1$ (at a given
value of $\theta_1$) to get
\begin{equation}
 \label{single-general}
 \frac{dw^{(s)}(\theta_1)}{d\theta_1}\propto\int d\theta
 \left|{\widetilde E}_p\left(\frac{\theta}{2}\right)\,{\rm sinc}\left[\frac{L}{2}\,
 \Delta_{z^\prime}(\theta;\,\theta_1)\right]\right|^2,
\end{equation}
where now
\begin{equation}
 \label{detuning-4}
 \Delta_{z^\prime}=\frac{k_p^{(0)}}{8}\Big[4n_p^\prime\theta+(\theta-2\theta_1)^2\Big].
\end{equation}
At $n_p^\prime\neq 0$ the sinc$^2$-function is not negligibly
small only in small vicinity of points where the detuning
$\Delta_{z^\prime}(\theta;\,\theta_1)$ turns zero. From the
condition $\Delta_{z^\prime}=0$ we get a quadratic equation in
$\theta$, solutions of which are given by
\begin{equation}
 \label{theta+-}
 \theta_{a,\,b}(\theta_1)=2\left[\,\theta_1-n_p^\prime\mp\sqrt{n_p^{\prime\;2}-2n_p^\prime\theta_1}\;\right].
\end{equation}
Only one of these two solutions ($\theta_a$) is small enough for a
region around this point to give a non-zero contribution to the
integral over $\theta$ in Eq. (\ref{single-general}). Near the
point $\theta_a$ the detuning $\Delta_{z^\prime}$ can be
approximated by a linear function of $\theta$
\begin{equation}
 \label{detuning-5}
 \Delta_{z^\prime}\approx\frac{k_p^{(0)}}{2}
 \sqrt{n_p^{\prime\;2}-2n_p^\prime\theta_1}\;(\theta-\theta_a)
\end{equation}
and the sinc$^2$ function can be approximated by the delta
function
\begin{equation}
 \label{delta-function}
 {\rm
 sinc}^2\left[\frac{L}{2}\Delta_{z^\prime}\right]
 \propto\frac{\delta(\theta-\theta_a)}{\sqrt{n_p^{\prime\;2}-2n_p^\prime\theta_1}}
\end{equation}
to give finally the following simple expression for the
single-particle momentum distribution of photons
\begin{equation}
 \label{single-final}
 \frac{dw^{(s)}_\|(\theta_1)}{d\theta_1}\propto
 \frac{\left|E_p\left[\displaystyle\frac{\theta_a(\theta_1)}{2}\right]\right|^2}
 {\sqrt{n_p^{\prime\;2}-2n_p^\prime\theta_1}}
\end{equation}
with $\theta_a(\theta_1)$ given by Eq. (\ref{theta+-}).

\begin{figure}[h]
\centering\includegraphics[width=8.5cm]{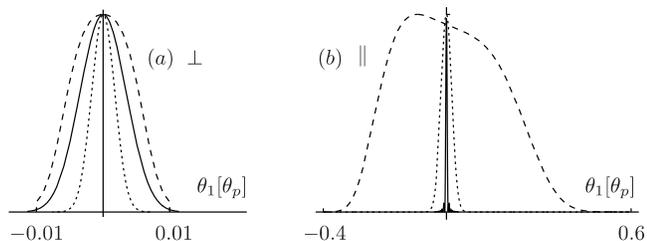}
\caption{{\protect\footnotesize {Coincidence (solid lines) and
single-particle (dashed lines) photon distributions,
$dw^{(c)}/d\theta_1$ (at $\theta_2=0$) and $dw^{(s)}/d\theta_1$,
calculated for the cases $(a)$ $n_p^\prime=0$ and
$n_p^\prime=-1.436$. The dotted-line curves show the pump
intensity in its dependence on $\theta_p$. All curves are
normalized by one at their maxima.}}} \label{Fig3}
\end{figure}

The coincidence (solid lines) and single-particle (dashed lines)
photon momentum probability distributions are plotted in Fig.
\ref{Fig3} for two cases, $(a)\;n_p^\prime=0$ ($\perp$-geometry)
and $(b)\;n_p^\prime\neq 0$ ($\|$-geometry). In both cases the
dotted-line curves describe the pump intensity $|{\widetilde
E}_p(\theta_p)|^2$

\noindent taken in the Gaussian form

\begin{equation}
 \label{Gaussian}
 |{\widetilde
 E}_p(\theta_p)|^2\propto\exp\left[-\frac{4\ln{(2)}\,\theta_p^2}{\alpha^2}\right]
\end{equation}
with $\alpha=0.004114$.

In the case $n_p^\prime=0$ ($\perp$-geometry) integration in Eq.
(\ref{single-general}) is performed numerically, whereas in the case
$n_p^\prime\neq 0$ ($\|$-geometry) we used the analytical expression
of Eq. (\ref{single-final}). On the other hand, as in the case
$n_p^\prime=0$ the sinc-function (Fig. \ref{Fig3}$(a)$) is wider
than the pump, the coincidence curve in the $\perp$-geometry is
close to the pump angular distribution curve in its dependence on
$(\theta_1+\theta_2)/2$ at a given $\theta_2$. In particular, at
$\theta_2=0$ the coincidence momentum probability distribution in
the $\|$-geometry is given by
\begin{equation}
 \label{coinc-perp}
 \frac{dw^{(c)}_\perp}{d\theta_1}\approx\left|E_p\left(\frac{\theta_1}{2}\right)\right|^2.
\end{equation}
Validity of this equation is pretty well confirmed by the
calculated coincidence and pump ($|{\widetilde E}_p(\theta_p)|^2$)
curves in Fig. \ref{Fig3}$(a)$: the solid-line curve is twice
wider than the dotted-line one.

As a whole, for the taken laser and crystal parameters, the case
$n_p^\prime=0$ ($\perp$-geometry) does not provide conditions for
observing high entanglement of the arising biphoton state. Indeed,
the coincidence and single-particle distributions are seen in Fig.
\ref{Fig3}$(a)$ to be rather close to each other. Their widths are
equal to $\Delta\theta_{1\,\perp}^{(c)}\approx 8$ mrad and
$\Delta\theta_{1\,\perp}^{(s)}\approx 12$  mrad, which corresponds
to the width ratio
$R_{k\,\perp}=\Delta\theta_{1\,\perp}^{(s)}/\Delta\theta_{1\,\perp}^{(c)}\approx
1.5$. In reality the width $\Delta\theta_1^{(s)}$ and the ratio
$R_{k\,\perp}$ can be slightly higher than estimated here because of
the non-collinear phase matching processes to be discussed below.
But this increase is not too high, and the main conclusion remains
valid: the $\perp$-geometry does not provide conditions for
observing high entanglement that can be accumulated in the biphoton
state under consideration.

This high entanglement can be seen most successfully in the
$\|$-geometry. The corresponding coincidence and single-particle
momentum distributions are shown for this case in Fig.
\ref{Fig3}$(b)$. As in the $\|$-geometry ($n_p^\prime\neq 0$) the
sinc-function in Eq. (\ref{coinc}) is much narrower than the pump,
the coincidence angular distribution is identical in this case to
the narrow peak of the sinc$^2$-function located near $\theta_1=0$
and described in Fig. \ref{Fig2}$(c)$:
\begin{eqnarray}
\frac{dw^{(c)}_\|(\theta_1)}{d\theta_1}={\rm
sinc}^2\left[\frac{Lk_p}{16}\,\left(4n_p^\prime\,\theta_1+\theta_1^2\right)\right].
\label{coinc-parll}
\end{eqnarray}
The width of this peak found above gives us the coincidence width of
the byphoton angular distribution in the $\|$-geometry: $\Delta
\theta_{1\,\|}^{(c)}=0.5$ mrad. The single-particle angular
distribution in the $\|$-geometry is determined by Eq.
(\ref{single-final}) and is shown in Fig. \ref{Fig3}$(b)$ in the
dashed-line curve. The width of this curve is equal to $\Delta
\theta_{1\,\|}^{(s)}=47.3$ mrad. The ratio of these widths is the
entanglement-parameter for measurements in the $\|$-geometry
\begin{equation}
 \label{R-paral}
 R_{k\,\|}=\frac{\Delta
\theta_{1\,\|}^{(s)}}{\Delta
\theta_{1\,\|}^{(c)}}=\frac{47.3}{0.5}=94.6\gg 1\,.
\end{equation}
In contrast to the earlier considered case of the $\perp$-geometry,
the $\|$-geometry of measurements does provide conditions for
observing very high degree of entanglement accumulated in the same
biphoton state and not seen so well in other geometries of
measurements.

Related to the estimate of the entanglement parameter $R_{k\,\|}$
(\ref{R-paral}), the two new specific effects predicted for the
$\|$-geometry and differing it from the traditionally considered
$\perp$-geometry are: (1) a very strong narrowing of the
coincidence and (2) broadening of the single-particle distribution
curves. The narrowing effect is very strong indeed: the width
$\Delta\theta_{1\,\|}^{(c)}$ is 16 times smaller than the same
width $\Delta\theta_{1\,\perp}^{(c)}$ occurring in the
perpendicular geometry, and 8 times smaller than the pump angular
width $\alpha$. This shows, in particular, that the coincidence
angular distribution appears to be narrower than that of the pump
(contrary to traditional expectations based on consideration of
the $\perp$-geometry scheme). Explanation of this effect is
related to drastic changes in the structure of the sinc-function
in Eqs. (\ref{new-via-theta}) and (\ref{coinc}) (see  Fig.
\ref{Fig2}), which occur when we move from the $\perp$- to
$\|$-geometry.

\begin{figure}[h]
\centering\includegraphics[width=8.5cm]{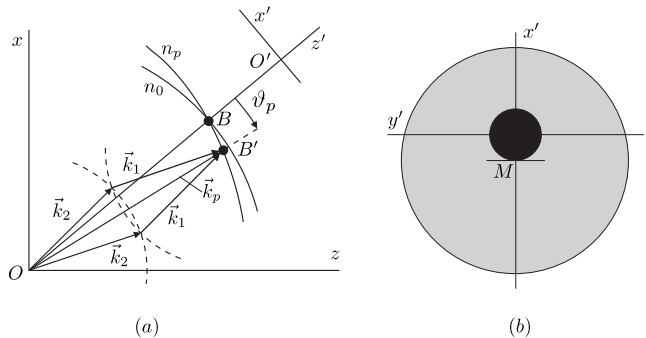}
\caption{{\protect\footnotesize {(a) Diagram of pump and emitted
photon wave vectors under the conditions of non-collinear phase
matching; dashed lines are arcs of circles with radii $n_o/2$ and
centers at the points $O$ and $B'$, wave vectors are in units of
$\omega_p/c$. (b) Angular distributions of photons emitted in the
regime of non-collinear phase matching (grey) and of the pump
(black), a view along the $-Oz^\prime$ direction, $M$ is the point
of the maximal achievable non-collinear phase matching.}}}
\label{Fig4}
\end{figure}

As for broadening of the single particle angular biphoton
distribution in the $\|$-geometry, it has a rather simple
qualitative explanation in terms of the non-collinear phase matching
down-conversion processes. Their origin is illustrated by two
pictures of Fig. \ref{Fig4}. The first picture $[4(a)]$ is the same
projection of the refractive index
surfaces as in Fig. \ref{Fig1}$(c)$. But now the pump wave vector
${\vec k}_p$ is taken slightly deflected from the laser axis
direction $Oz^\prime$ (upon the angle $\vartheta_p$). As ${\vec
k}_p(\vartheta_p)$ is shorter than ${\vec k}_p(0)$, its ending point
$B^\prime$ cannot be reached by two collinear wave vectors ${\vec
k}_1$ and ${\vec k}_2$, and these vectors must be non-collinear both
to each other and to ${\vec k}_p(\vartheta_p)$. As it's seen from
Fig. \ref{Fig4}$(a)$, the wave vectors ${\vec k}_1$ and ${\vec k}_2$
providing fulfilment of the exact phase matching condition are
emitted under angles (with respect to the $Oz^\prime$ axis)
exceeding $\vartheta_p\,$. This explains why the non-collinear phase
matching emission broadens the single-particle probability density
curves.

By rotating the wave-vector diagram of Fig. \ref{Fig4}$(a)$ around
the direction of the pump wave vector ${\vec k}_p$ we get cones
along which the non-collinear-emitted photons can propagate. The
cone opening angle changes with changing $\vartheta_p$ by forming in
such a way the total emission area. Schematically this area of
emission (in the angular space) is shown in Fig. \ref{Fig4}$(b)$
(shaded in grey). This area is seen to be wider than that of the
pump (shaded in black) and it's seen to be wide both in the
$O^\prime x^\prime$ and $O^\prime y^\prime$ directions. The last
case corresponds to detection of photons in the $\perp$-plane.
Hence, as said above, emission of photons in the non-collinear phase
matching regime can make the single-particle probability density
distribution somewhat wider than that described by the dashed-line
curve in Fig. \ref{Fig3}$(a)$.

The point $M$ in Fig. \ref{Fig4}$(b)$ indicates the maximal
achievable angle $\vartheta_p$ evaluated, e.g., at a half-width of
the pump angular distribution, $\vartheta_p=\alpha/2$. This value of
$\vartheta_p$ provides the maximal emission cone opening angle,
which is characterized by the circle in Fig. \ref{Fig4}$(b)$
bordering the grey area and having its center at $M$. It's seen in
Fig. \ref{Fig4}$(b)$ that the emission are area is asymmetric with
respect to the point $O^\prime$ ($x^\prime$ and $z^\prime$ line
crossing) in the $O^\prime x^\prime$ direction. This agrees with the
structure of the single-particle probability density distribution in
the case of detection in the $\|$-plane [the dashed-line curve of
Fig. \ref{Fig3}$(b)$]. The reason of asymmetry is evident:
deflection of the pump wave vector ${\vec k}_p$ from the $Oz^\prime$
axis to the direction opposite to that shown in Fig. \ref{Fig4}$(a)$
makes ${\vec k}_p$ longer than ${\vec k}_p(0)$. For such wave
vectors, in contrast to the case considered above, the exact phase
matching condition cannot be fulfilled at all and, hence, such
deflections give almost zero contribution to the integral in Eq.
(\ref{single-final}) determining the single-particle probability
density distribution.

Note that the role of photon emission in the regime of the
non-collinear phase matching can be diminished by special profiling
the pump angular distribution. In terms of Fig. \ref{Fig1}
geometries, if the pump angular distribution is made narrow in the
$O^\prime x^\prime$ direction but, still, remains wide enough in the
$O^\prime y^\prime$ direction, then the point $M$ in Fig.
\ref{Fig4}$(b)$ approaches $O^\prime$ and the radius of the grey
area decreases, thus making single-particle angular distributions
narrower. In the extreme limit of the pump angular distribution very
narrow in the $O^\prime x^\prime$ direction, in the case of photon
detection in the $\perp$-geometry we return to the results described
by the curves of Fig. \ref{Fig3}$(a)$. Such an asymmetric profiling
of the pump angular distribution can be provided by focusing with a
cylindrical lens or by a slit.

So, the used above simplification that the essential pump wave
vectors ${\vec k}_p$ belong to the same plane as ${\vec k}_1$ and
${\vec k}_2$ can be not too good for describing the
single-particle angular photon distribution in the
$\perp$-geometry. But this simplification does not affect both
coincidence distributions at arbitrary orientation of the
observation plane $(z^\prime\xi)$ and any distributions in the
$\|$-geometry.

\subsection{Coordinate representation}

The biphoton wave function in the coordinate representation can be
obtained from the momentum-representation wave function  by means
of a double Fourier transformation
\begin{eqnarray}
 \notag\Psi_{\rm coord}(\xi_1,\,\xi_2)=\int dk_{1\,\xi}\,dk_{2\,\xi}\,
 \Psi_{\rm mom}(k_{1\,\xi},\,k_{2\,\xi})\\
 \times\exp\big[i(\xi_1 k_{1\,\xi}+\xi_2 k_{2\,\xi})\big],\quad\quad\quad
 \label{coord-wf}
\end{eqnarray}
where $\xi_1$ and $\xi_2$ are coordinates along the observation
direction $O^\prime\xi$ for photons (1) and (2) and $\Psi_{\rm
mom}(k_{1\,\xi},\,k_{2\,\xi})$ is the same function as
$\Psi(k_{1\,\xi},\,k_{2\,\xi})$ of Eq. (\ref{New-formula}).
Integrations in Eq. (\ref{coord-wf}) can be carried out partially
in a rather simple way only in the case of a sufficiently large
refractive index derivative. So let us consider only the
$\|$-geometry when the detection direction $O^\prime\xi$ coincides
with $O^\prime x^\prime$. The integration variables
$k_{1\,x^\prime}$ and $k_{2\,x^\prime}$ can be substituted by the
angular variables
$\theta_p=(k_{1\,x^\prime}+k_{2\,x^\prime})/k_p^{(0)}$ and
$\theta=(k_{1\,x^\prime}-k_{2\,x^\prime})/k_p^{(0)}$. In these
variables the sinc-function in (\ref{New-formula}) takes the form
convenient for its subsequent approximation by the
$\delta$-function:
\begin{equation}
 \label{sinc-delta}
 {\rm sinc}\left[\frac{Lk_p^{(0)}}{2n_o}\left(n_p^\prime\theta_p+\frac{\theta^2}{2}\right)\right]
 \propto\delta\left(\theta_p+\frac{\theta^2}{2n_p^\prime}\right)\,.
\end{equation}
In this approximation and with the pump amplitude ${\widetilde E}_p$
taken in the Gaussian form [Eq. (\ref{Gaussian})], Eq.
(\ref{coord-wf}) yields
\begin{eqnarray}
 \Psi_{\rm coord}(x^\prime_1,\,x^\prime_2)\propto \int d\theta
 \exp\left\{-\frac{\ln(2)\theta^4}{2\alpha^2n_p^{\prime\;2}}\right.\notag\\
 +i\,k_p^{(0)}\left.\left[-\frac{x^\prime_1+x^\prime_2}{4n_p^\prime}\,\theta^2+
 \frac{x^\prime_1-x^\prime_2}{2}\,\theta\right]\right\}\,.\label{coord-wf-final}
\end{eqnarray}
The $\delta$-function approximation of (\ref{sinc-delta}) and the
final expression of Eq. (\ref{coord-wf-final}) are valid  the
sinc-function on the left-hand side of Eq. (\ref{sinc-delta}) is
narrower than the pump ${\widetilde E}_p(\theta_p)$ of Eq.
(\ref{Gaussian}), which gives
\begin{equation}
 \label{condition}
 \frac{Lk_p^{(0)}|n_p^\prime|}{2n_0}\gg\frac{2}{\alpha}\,.
\end{equation}
With $L$, $\alpha$, and $k_p^{(0)}$ we use in this paper the left-
and right-hand sides of this inequality are equal to $1.1\times
10^4$ and 486, correspondingly, and the condition of Eq.
(\ref{condition}) is pretty well satisfied. Under the condition
(\ref{condition}) the coordinate biphoton wave function
(\ref{coord-wf-final}) does not depend on the crystal length $L$.
Its features are fully controlled by the pump angular divergence
$\alpha$.

\begin{figure}[h]
\centering\includegraphics[width=6cm]{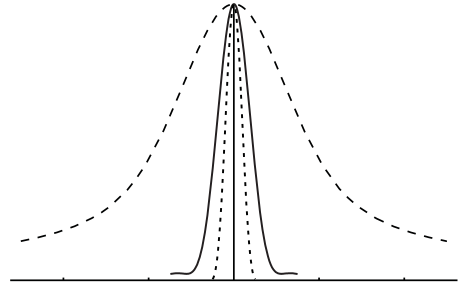}
\caption{{\protect\footnotesize {Coincidence probability density in
the coordinate representation in its dependence on $x_1^\prime
k_p^{(0)}/2$ (solid line); dependence of the bipartite probability
density on $(x_1+x_2)k_p^{(0)}/2$ at $x_1-x_2=0$ (dashed line) and
on $(x_1-x_2)k_p^{(0)}/2$ at $x_1+x_2=0$ (dotted line).}}}
\label{Fig5}
\end{figure}

The biphoton coincidence coordinate distribution
$dw^{(c)}/dx_1^\prime|_{x_2^\prime=0}\propto|\Psi_{\rm
coord}(x_1^\prime,\,0)|^2$ calculated with the help of Eq.
(\ref{coord-wf-final}) is plotted in Fig. \ref{Fig5} in its
dependence on

\noindent the dimensionless product $x_1^\prime k_p^{(0)}/2$ (the
solid-line curve). In the same picture the probability density
$|\Psi_{\rm coord}|^2(x_1^\prime,x_2^\prime)$ is plotted in its
dependence on $\xi_+\equiv(x_1^\prime+x_2^\prime)k_p^{(0)}/2$ at
$\xi_-\equiv (x_1^\prime-x_2^\prime)k_p^{(0)}/2=0$ (the
dashed-line curve) and on $\xi_-$ at $\xi_+=0$ (the dotted-line
curve). The widths of these curves are equal to $\Delta
x_1^{\prime\,(c)}k_p^{(0)}/2=88$, $\Delta\xi_+=356.4$, and
$\Delta\xi_-=44$. The distribution is seen to be much wider in the
$\xi_+$ than in $\xi_-$ direction.

We did not calculate (yet) the single-particle probability density
in the coordinate representation. But we can use the found width of
the coincidence coordinate distribution for evaluating the
EPR-related parameter of entanglement. Indeed, by taking into
account that $\Delta k_{1\,x^\prime}=\Delta\theta_1\times
k_p^{(0)}/2$, we get for the $\|$-geometry: $\Delta
k_{1\,x^\prime}^{(c)}\times\Delta x_1^{\prime\,(c)}=0.0005\times
88=0.0044$ and $C_{\rm EPR}=1/\Delta
k_{1\,x^\prime}^{(c)}\times\Delta x_1^{\prime\,(c)}=22.7$. By taking
into account that in the no-entanglement case for Gaussian functions
and for our definitions of widths as the half-height widths $C_{\rm
EPR}^{(0)}=(4\ln 2)^{-1}$, we find finally the the EPR related
parameter of entanglement for the $\|$-geometry is evaluated as
\begin{equation}
 \label{EPR}
 \frac{C_{\rm EPR}}{C_{\rm EPR}^{(0)}}=22.7\times 4\ln 2 = 63.
\end{equation}
Though less than 80, this parameter is also rather large and
indicates clearly that SPDC biphoton states under consideration are
rather highly entangled.

\section{Experiment}
\subsection{Experimental setup}

The experimental setup is shown on Fig.3. To generate the entangled
photons we use type I and
 $15$ mm-length lithium-iodate crystal pumped with a 5 mW cw-
helium-cadmium laser operating at $325$ nm. The correlated photons
generated via SPDC process with equal polarization and wavelength
$650$ nm are separated from the pump by dichroic mirror (DM).
Interference filters centered at $650$ nm with a bandwidth of $10$
nm are placed in each arm of Brown-Twiss scheme. To measure
coincidence and single-photon distributions in the transverse
momenta we use the lens with focal length $F=62$ cm. Two single
photon detectors are positioned in focal plane of the lens. Such
detector arrangement allows one to measure the momentum
distribution(s) by scanning one or both detectors along with
certain direction (see below). In the most of cases we fix
position of the first detector at the maximum of count rate and
scan another one to register both distributions as a function of
detector displacement. Since detector moves in the focal plane its
position ($x$) relates to the angular mismatch $(\theta)$ as $x=F
\tan\theta$.
\begin{figure}[h]
\centering\includegraphics[width=8cm]{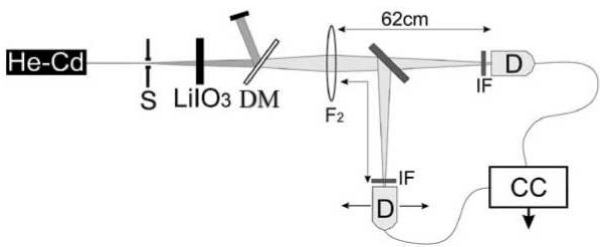}
\caption{{\protect\footnotesize {Experimental setup for measuring
single and coincidence probability distributions.}}} \label{Fig6}
\end{figure}

\subsection{Results and discussion}
The main idea behind performed experiment is to check formulas
(\ref{single-final}), (\ref{coinc-parll}) and (\ref{coinc-perp}) for
two geometries such that detectors are scanned (*) in the $\|$-plane
containing optical axis, when $O^\prime\xi=O^\prime x^\prime\in
(xz)$, and (**) in the $\perp$-plane, when $O^\prime\xi=O^\prime
y^\prime\|Oy$ (see Fig. \ref{Fig1}). One of the key parameters of
the theory described above is the pump angular width. Originally our
He-Cd laser had the angular width equal to 1.5 mrad. To see in
experiment how the angular distrubution of the pump affects upon
biphoton angular distributions, we have artificially and
anisotropically broadened the pump distribution in angles by
installing in front of the crystal a slit. As a result, the pump
angular distribution remains localized along the direction parallel
to the slit (at the same level of 1.5 mrad as was without a slit)
but it spreads in the orthogonal direction. The slit was measured to
be 40 mkm wide, which corresponds to the pump angular width in the
direction perpendicular to the slit equal to 4.1 mrad. Alternatively
we have used in some measurements a lens instead of a slit to
provide axially symmetric (isotropic) broadening of the pump up to
the same width of 4.1 mrad. Comaprison of results of such
measurements was used for evaluating the role of the pump angular
broadening for efficiency of emission processes arising under the
non-collinear phase matching conditions.

As mentioned above, in the experiment the detection plane was always
horizontal. The slit was installed always vertically to provide
angular broadening in the horizontal direction, whereas the crystal
optical axis could be lying either in the vertical or horizontal
planes. In terms of notations and definitions used above and
introduced in the picture of Fig. \ref{Fig1}$(a)$ this means that
our experimental measurements correspond to one of the following two
situations: (A) detection and pump angular broadening directions are
along the $O^\prime x^\prime$ axis, i.e., in the $\|$-plane or in
the plane containing laser and optical axes and (B) both detection
and pump angular broadening directions are along the $O^\prime
y^\prime$ axis, i.e., in the $\perp$-plane, and these directions are
perpendicular to the plane containing laser and optical axes. In the
case when lens is used for the pump broadening it provides equal
angular broadening in $O^\prime x^\prime$ and and $O^\prime
y^\prime$, independently of the detection direction. As in the case
$B$ the slit-broadened pump angular distribution is narrow in the
$O^\prime x^\prime$ direction, we expect that under these conditions
non-collinear phase matching processes will be suppressed [see the
explanantion in Fig. \ref{Fig4}$(b)$] and the single-particle
angular distribution in the direction $O^\prime y^\prime$ will be
narrower than in the case of lens angular broadening.

Another key parameter is angular derivative of the pump
extraordinary refractive index $n_p^\prime$ near the exact phase
matching direction $Oz^\prime$. The table 1 shows this value for
different crystals available for producing photon pairs. It shows
that Lithium Iodate is the best candidate because in this crystal
the derivative $|n_p^\prime|$ takes a rather large value. Besides,
the effect of high entanglement anisotropy discussed in the paper
depends strongly on the product $Ln_p^\prime$. So, the second reason
why we chose this crystal is that the sample of lithium iodate can
be made rather long.
\begin{table}[!ht]
\begin{tabular}{|c||c|c|}\hline
Crystal&phase matching angle $\varphi_0$ (deg.)&$n_p^\prime$\\
\hline
$LBO$&$51.47$&$-0.0270$\\
\hline
$KDP$&$54.33$&$-0.0395$\\
\hline
$BBO$&$36.44$&$-0.1175$\\
\hline
$LiIO_3$&$5.97$&$-0.1409$\\
\hline
\end{tabular}
\caption{Angular anisotropy parameter $n_p^\prime$ for different
non-linear crystals}
\end{table}

As it follows from theory developed above, at sufficiently high
values of $|n_p^\prime|$ the coincidence angular distribution of
biphotons does not depend at all on the divergence of the pump [see
Eq. (\ref{coinc-parll})]. Moreover, being determined completely by
properties of the crystal and, in particular, by anisotropy of the
refractive index, the width of the coincidence distribution can be
done even narrower than the pump angular width because. At the same
time the width of the single-photon distributions grows up with the
pump broadening. Both narrowing of the coincidence and broadening of
the single-particle angular distributions are factors resulting in a
growing degree of entanglement.

The pictures of Fig. \ref{Fig7} show two sets of angular
distributions received in (a) single-particle and (b) coincidences
measurements, which are plotted together for different geometries,
with a slit broadened
\begin{figure}[h]
\centering\includegraphics[width=8cm]{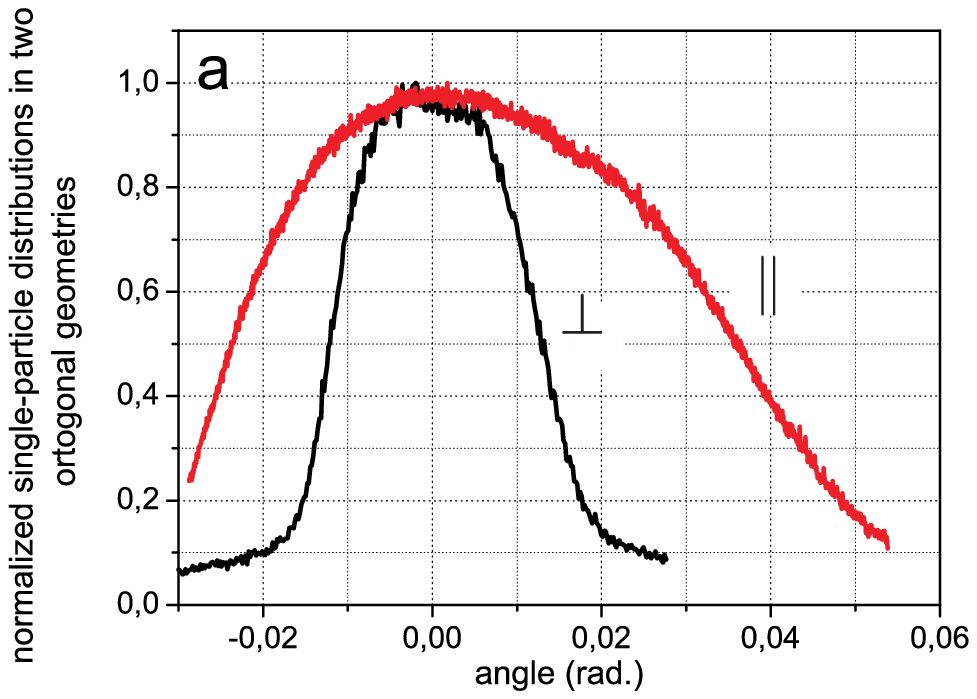}
\end{figure}
\begin{figure}[h]
\centering\includegraphics[width=8cm]{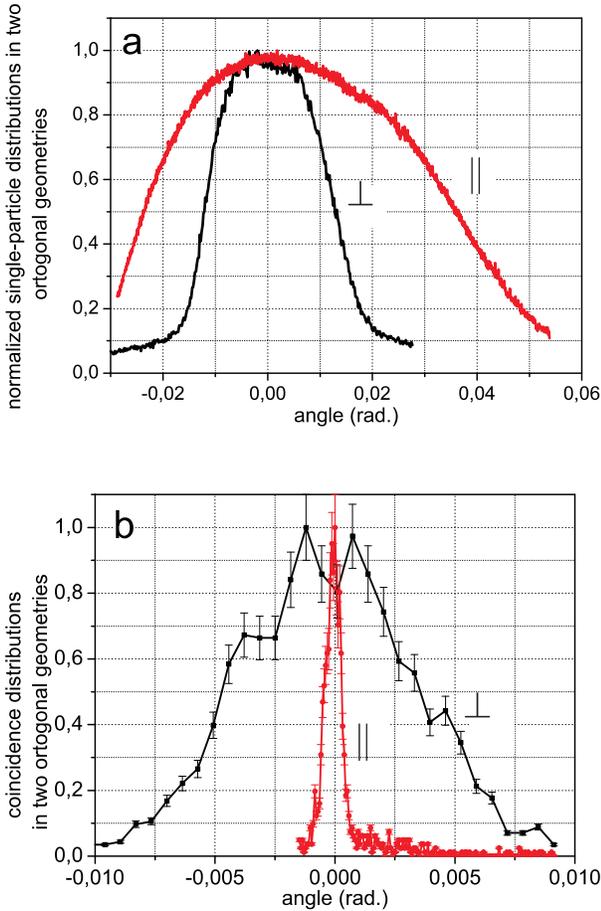}
\caption{{\protect\footnotesize {Angular distributions of (a) single
and (b) coincidences counts for two orthogonal geometries ($\perp$)
and ($\|$)}.}} \label{Fig7}
\end{figure}
pump beam. These pictures illustrate clearly that in the
$\|$-geometry coincidence distribution becomes narrower whereas the
single-particle distribution broadens in comparison with the
$\perp$-geometry. The corresponding  ratios are $\Delta
k_{\perp}^{(c)}/\Delta k_{\|}^{(c)}$=11 for coincidences and $\Delta
k_{\perp}^{(s)}/\Delta k_{\|}^{(s)}$=0.41 for singles. For
comparison, the corresponding theoretical estimates are $\Delta
k_{\perp}^{(c)}/\Delta k_{\|}^{(c)}\approx 16$ and $\Delta
k_{\perp}^{(s)}/\Delta k_{\|}^{(s)}\approx 0.25$. The difference
between theoretical and experimental results is not too high, though
quite visible. Probably, it has different origin for coincidence and
single width ratios. In the case of coincidence counts, we think
that the main reason for a difference between the theoretical and
experimental width ratios is related to some external factors owing
to which the experimentally measured coincidence width is larger
than the theoretical estimate (0.75 mrad in experiment compared to
0.5 mrad in theory). As for the single-particle width, probably in
experiment the corresponding curve in the $\perp$-geometry did
experience some broadening arising from the non-collinear phase
matching emission processes, in spite of the missing slit-broadening
of the pump angular distribution in the $O^\prime x^\prime$
direction. In contrast to this, in theory presented above in the
$\perp$-geometry case the non-collinear phase matching emission
processes were completely excluded because only pump photons with
wave vectors belonging to the observation plane were taken into
account. Hence, compared to the real experimental situation, the
given above theoretical derivation of the single-particle angular
distribution in the $\perp$-geometry artificially narrows the
corresponding curve, and this is the reason for the observed
discrepancy between theory and experiment. But let us emphasize once
again: the discrepancy is not too high, and the degree of closeness
of theoretical and experimental results can be considered as quite
satisfactory.

\begin{figure}[h]
\centering\includegraphics[width=8cm]{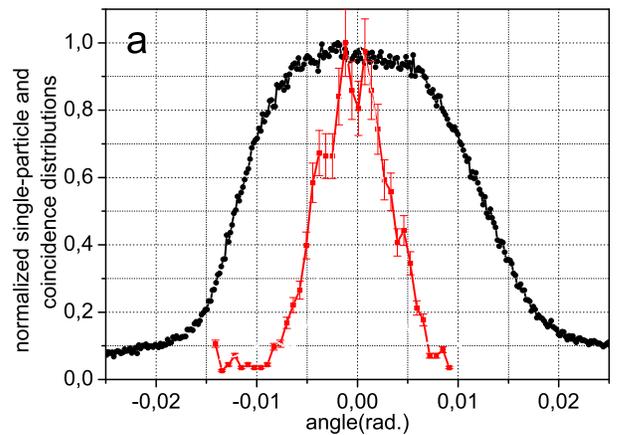}
\caption{{\protect\footnotesize {Experimental results: normalized
single-particle and coincidence distributions for $\perp$-geometry.
Curves with errors are the coincidence. The pump is broadened by a
slit.}}} \label{Fig8}
\end{figure}

\begin{figure}[h]
\centering\includegraphics[width=8cm]{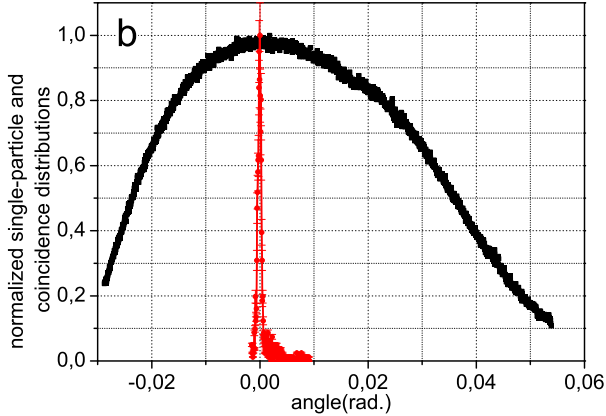}
\caption{{\protect\footnotesize {Experimental results: normalized
single-particle and coincidence distributions for $\|$-geometry.
Curves with errors are the coincidence. The pump is broadened by a
slit.}}} \label{Fig9}
\end{figure}

Figs. \ref{Fig8} and \ref{Fig9} present the same experimental
results only differently regrouped, which allows one to evaluate the
experimentally found degree of entanglement.

Fig. \ref{Fig8} corresponds to the case when detector is scanning in
the plane perpendicular to the optical axis ($\perp$-geometry), and
the slit broadens the pump angular distribution in the same
direction. The width of the single-

\noindent particle distribution is 25 mrad whereas the width of the
coincidence one is 8.4 mrad, i.e., twice wider than the width of the
pump [in accordance with Eq. (\ref{coinc-perp})]. The ratio
$\frac{\Delta k^{(s)}_{1\,\perp}}{\Delta k^c{(s)}_{1\,\perp}}=3$, so
the degree of entanglement is not very high. Note, that in the
above-presented theory we got for this ratio the twice smaller
value, 1.5. The reason is the same as discussed above: in theory the
single-particle angular distribution is artificially narrowed
because all non-collinear phase matching processes are completely
excluded from the consideration.

The results occurring in the case when both scanning and pump
broadening occur in the plane containing the crystal optical axis
($\|$-geometry) are shown in Fig. \ref{Fig9}. Here the widths of
single- and coincidence distributions

\noindent are 60 mrad and 0.75 mrad correspondingly. Their ratio is
80, which is much greater than in previous case though is somewhat
less than the corresponding theoretical estimate $R_{k\,\|}=94.6$
(\ref{R-paral}). The difference can be attributed only to some
external factors affecting experiment. In the case of the
$\|$-geometry the above-discussed non-collinear phase matching
processes are completely taken into account in the theoretical
derivation, and the explanation of the theory-experiment
discrepancies given above for the $\perp$-geometry does not work for
the $\|$-geometry.

Fig. \ref{Fig10} presents similar graphs but measured for the case
when the laser pump is broadened with a spherical lens. This means
that the pump broadening affects angular distributions in both
directions and so, it is impossible to distinguish its contribution
to spreading in $\|$- or $\perp$-planes. According to the picture of
Fig.\ref{Fig4}$(b)$, the single-particle distribution measured in
$\perp$-plane must be wider when the pump angular distribution is
broadened by a lens compared to the case of broadening by a slit.
This expectation is confirmed by measurements which give (for the
$\perp$-geometry) $\Delta k_{(\rm lens)}^{(s)}/\Delta k_{(\rm
slit)}^{(s)}$=0.075/0.025=3. However, another observed effect
remains unexplained: for the same $\perp$-geometry the coincidence
width in the case of lens-broadening appears to be smaller than in
the case slit-broadening, $\Delta k_{(\rm lens)}^{(c)}/\Delta
k_{(\rm slit)}^{(c)}$=0.0031/0.0084=0.37. Note also that in the case
of lens broadening the values of the entanglement parameter $R_k$
appear to be somewhat smaller than in the case of slit broadening in
both $\perp$- and $\|$-geometries: in the case of lens broadening
$R_{k\,\perp}=2.33$ and $R_{k\,\|}=67$. But the main qualitative
conclusions remain the same as earlier: at chosen values of the
parameters almost no entanglement can be seen in the
$\perp$-geometry and a very high entanglement can be and was
observed in the $\|$-geometry.

\begin{figure}[h]
\centering\includegraphics[width=8cm]{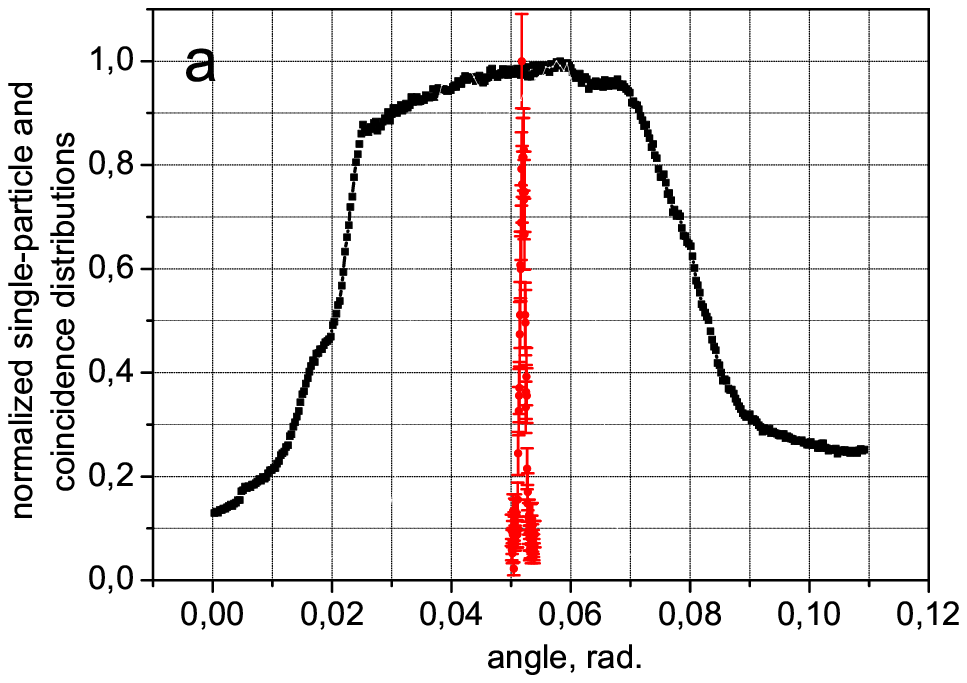}
\end{figure}
\begin{figure}[h]
\centering\includegraphics[width=8cm]{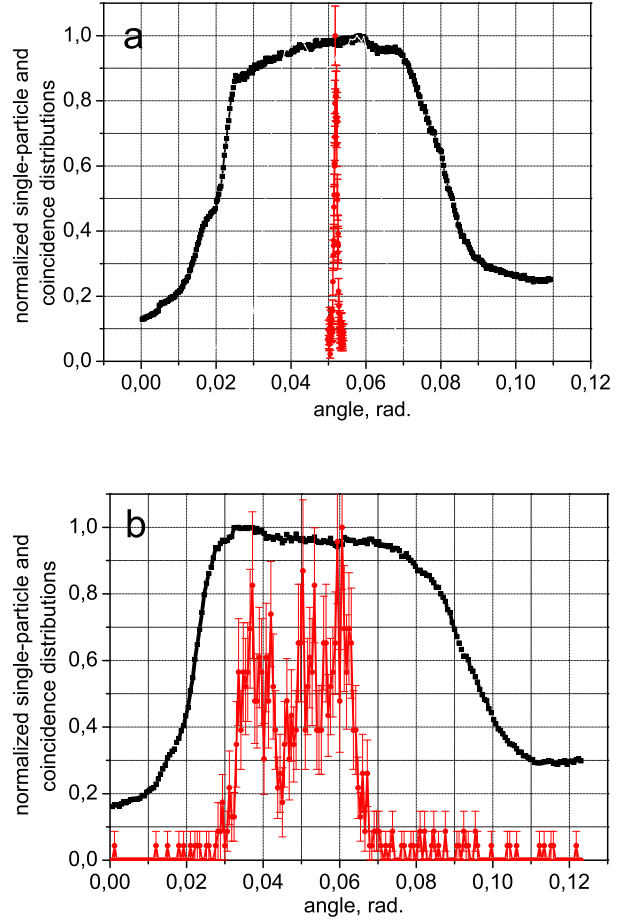}
\caption{{\protect\footnotesize {Experimental results: normalized
single-particle and coincidence distributions for (a) $\|$-   and
(b) $\perp$-geometries. Curves with errors are the coincidence.
The pump is broadened by a lens.}}} \label{Fig10}
\end{figure}

In conclusion, we have shown both theoretically and experimentally
that there are two different schemes of observing SPDC biphoton wave
packets. In the traditional and alternative schemes the detector
scanning is assumed to be performed in the planes, correspondingly,
perpendicular and parallel to the plane containing optical and laser
axes. Owing to the anisotropy of the crystal refractive index for
the extraordinary wave, a structure of the coincidence and
single-particle biphoton wave packets observable in these two
schemes are significantly different. In the alternative scheme,
coincidence wave packets demonstrate a very strong narrowing
compared to the traditional scheme, whereas the single-particle wave
packets broaden. All this results in a very high degree of
entanglement that can be observed in the alternative scheme of
observations, and cannot be seen in the traditional scheme. The
entanglement parameter determined as the ratio of the single to
coincidence wave packet widths appears to be as high as about 100.
The degree of the coincidence wave packet narrowing is shown to be
so strong that in the alternative scheme of observations the
coincidence wave packet appears to be much narrower than the angular
distribution of the pump.

\section{Acknowledgments}

This work was supported in part by the Russian Foundation for
Basic Research (projects 05-02-16469 and 06-02-16769), the RF
President's Grant MK1283.2005.2, the Leading Russian Scientific
Schools (project 4586.2006.2), and by the US Army International
Technology Center - Atlantic, grant RUE1-1616-MO-06.

\end{document}